\documentclass{llncs}
\usepackage{graphicx}
\usepackage{amssymb} 
\usepackage{amsmath, amsfonts}

\usepackage{hyperref}

\usepackage{subcaption}

\usepackage[inline]{enumitem} 

\usepackage{caption}
\usepackage{subcaption}

\usepackage{amsthm}

\usepackage{tikz}

\usetikzlibrary{arrows,chains,fit}
\usetikzlibrary{automata,positioning,shapes}

\usepackage{graphicx}

\usepackage{listings}

\usepackage{makecell}

\usepackage{lmodern}

\newcommand{\comment}[1]{}

\newcommand{\state}{\mathit{State}}
\newcommand{\stmt}{\mathit{Stmt}}

\newcommand{\set}[1]{\left\{ #1 \right\}}

\newcommand{\denot}[1]{\left[\!\left[#1\right]\!\right]}

\newcommand{\toolname}[1]{\textsc{#1}}
\newcommand{\nuxmv}{\toolname{nuXmv}}

\newcommand{\tlc}{\toolname{TLC}}
\newcommand{\tlaplus}{\mbox{\toolname{TLA}}^+}

\newcommand{\idaction}{\mbox{$\mathit{id}$}}

\newcommand{\bexpr}{\mathit{BExpr}}

\newcommand{\loc}[1]{#1^{l}}
\newcommand{\glob}[1]{#1^{g}}

\newcommand{\sloc}{\loc{s}}
\newcommand{\sglob}{\glob{s}}
\newcommand{\locinit}{\mathit{init}^l}

\newcommand{\nuxmvcmd}[1]{\mathit{#1}}

\newcommand{\nunext}[1]{\nuxmvcmd{next}(#1)}

\definecolor{mGreen}{rgb}{0,0.6,0}
\definecolor{mGray}{rgb}{0.5,0.5,0.5}
\definecolor{mPurple}{rgb}{0.58,0,0.82}
\definecolor{backgroundColour}{rgb}{0.95,0.95,0.92}
\definecolor{bgColC}{rgb}{0.95,0.95,0.92}
\definecolor{nodeGray}{rgb}{0.98,0.98,0.98}

\lstdefinestyle{CStyle}{
    backgroundcolor=\color{backgroundColour},   
    commentstyle=\color{mGray},
    keywordstyle=\color{magenta},
    numberstyle=\tiny\color{mGray},
    stringstyle=\color{mPurple},
    basicstyle=\scriptsize\ttfamily,
    breakatwhitespace=false,         
    breaklines=true,                 
    captionpos=b,                    
    keepspaces=true,                 
    numbers=left,                    
    numbersep=5pt,                  
    showspaces=false,                
    showstringspaces=false,
    showtabs=false,                  
    tabsize=2,
    language=C
}

\lstdefinestyle{CStyle2}{
  backgroundcolor=\color{bgColC},
  commentstyle=\color{mGray},
  keywordstyle=\color{magenta},
  numberstyle=\tiny\color{mGray},
  stringstyle=\color{mPurple},
  basicstyle=\scriptsize,
  breakatwhitespace=false,
  breaklines=true,
  captionpos=b,
  keepspaces=true,
  numbers=left,
  numbersep=5pt,
  showspaces=false,
  showstringspaces=false,
  showtabs=false,
  tabsize=2,
  language=C
}

\lstdefinestyle{ACSLStyle}{
    style=CStyle,
    keywordstyle=[1]\color{magenta},
    keywordstyle=[2]\ttfamily\bfseries\color{mGreen},    
    morekeywords = [2]{ensures,requires,assigns},
    moredelim =*[s][\color{mGreen}]{/*@}{*/}
}

\lstdefinelanguage{TLA}{
    alsoletter={\\},
  backgroundcolor=\color{backgroundColour},
  commentstyle=\color{mGray},
  keywordstyle=\color{blue},
  numberstyle=\tiny\color{mGray},
  stringstyle=\scriptsize\ttfamily,
  basicstyle=\scriptsize\ttfamily,
  keywords={EXTENDS, CONSTANT, VARIABLES, MODULE, Spec, Init, M},
  breakatwhitespace=false,
  breaklines=true,
  captionpos=b,
  keepspaces=true,
  numbers=left,
  numbersep=5pt,
  showspaces=false,
  showstringspaces=false,
  showtabs=false,
  tabsize=2,
  upquote=true,
  morecomment=[l]{\\*}
}

\lstdefinelanguage{SMV}{
    alsoletter={\\},
  backgroundcolor=\color{backgroundColour},
  commentstyle=\color{mGray},
  keywordstyle=\color{blue},
  numberstyle=\tiny\color{mGray},
  stringstyle=\scriptsize\ttfamily,
  basicstyle=\scriptsize\ttfamily,
  keywords={DEFINE, INIT, TRANS},
  breakatwhitespace=false,
  breaklines=true,
  captionpos=b,
  keepspaces=true,
  numbers=left,
  numbersep=5pt,
  showspaces=false,
  showstringspaces=false,
  showtabs=false,
  tabsize=2,
  upquote=true,
  morecomment=[l]{--}
}

\newcommand{\Stee}{\textsc{Stee}}

\tikzstyle{arrow2} = [thick,<->,>=stealth]

\tikzset{every initial by arrow/.style={->}
}
\tikzset{initial text={}}

\tikzset{every label/.style={align=left}}

\tikzstyle{nodeone} = [rounded corners, text width=2.3cm, minimum height=1.8cm,text centered, draw=black, fill = nodeGray, font = \scriptsize]

\tikzset{flownode/.style={ 
        minimum size=2mm, 
        minimum width=2mm,
        rectangle, 
        draw=black,
        fill=black
    }
}

\tikzset{regnode/.style={ 
        flownode,
        initial above, 
        label={[font=\scriptsize,text=blue]right:{#1}},
    },
    regnode/.default={$\idaction$}
}

\tikzset{entrynode/.style={ 
        flownode,
        initial above, 
        label={[font=\scriptsize,text=blue]right:#1}
    }
}

\tikzset{retnode/.style ={ 
        flownode, 
        label={[font=\scriptsize,text=blue]right:#1}
    }
}

\tikzset{labelnode/.style={ 
        font=\scriptsize
    }
}

\tikzset{prognode/.style={ 
        minimum size=2mm, 
        minimum width=2mm,
        rectangle, 
        draw=black,
    }
}

\tikzset{pregnode/.style={ 
        prognode,
    },
}

\tikzset{pentrynode/.style={ 
        ellipse,draw=black
    }
}

\tikzset{pretnode/.style ={ 
        pregnode,
        append after command={
        (\tikzlastnode) -- (0,0)
    }
}
}
\newcommand{\pretnode}[3]{
    \node[pregnode,#3] (#1) {#2};
    \draw[->, bend left] (#1.south) to ++(-0.5em,-1em);
}

\title{Contract Based Program Models \\ for 
Software Model Checking\thanks{This work has been funded by the FFI Programme of the Swedish Governmental Agency for Innovation Systems (VINNOVA) as the AVerT2 project 2021-02519.}
}
\author{Jesper Amilon \and Dilian Gurov}
\authorrunning{Amilon et al.}
\institute{KTH Royal Institute of Technology, Stockholm, Sweden \\ \email{\{jamilon,dilian\}@kth.se}}

\begin{document}

\maketitle              

\begin{abstract}
Model checking temporal properties of software is algorithmically hard. To be practically feasible, it usually requires the creation of simpler, abstract models of the software, over which the properties are checked. However, creating suitable abstractions is another difficult problem.
We argue that such abstract models can be obtained with little effort, when the state transformation properties of the software components have already been deductively verified. As a concrete, language-independent representation of such abstractions we propose the use of \emph{flow graphs}, a formalism previously developed for the purposes of compositional model checking.
In this paper, we describe how we envisage the work flow and tool chain to support the proposed verification approach in the context of embedded, safety-critical software written in~C. 
\end{abstract}


\section{Introduction}
\label{sec:introduction}

In previous work~\cite{ami-lid-gur-22-isola}, we introduced a workflow for software model checking of C programs, the key point of which being that we utilised deductive verification of contracts for creating abstract models of C programs. When creating the models for model checking, we first replaced certain components with a Hoare-style contract, thus viewing those components purely as state-transformers, ignoring their intermediate states. The abstraction is justified by deductively verifying that the components indeed satisfy their contracts. 

In~\cite{ami-lid-gur-22-isola}, we used as tool chain the WP plugin of Frama-C~\cite{wpplugin} for deductive verification, and \tlc{}~\cite{yu-et-al-99-tlc} as model checker.
The translation from a program with deductively verified contracts to a $\tlaplus$ specification was partially performed with the support of the $\mathrm{C2TLA^+}$~\cite{methni-et-al-14} tool. However, much of the translation process was carried out
manually, in an ad hoc fashion. 
In this paper, we propose an explicit model of the artefact corresponding to a given program with deductively verified contracts. This model, called \emph{flow graph},  abstracts from the code, keeping just the information needed to verify temporal properties of the original program. The model is independent of the actual model checking technique, and allows automated translations to the input language of the desired model checker, be it \tlc{} or \nuxmv{}. This paper is an extended version of our previous work~\cite{ami-gur-24-isola}.
The model is adapted from previous work on modular model checking~\cite{sol-gur-16-scp}.

\begin{figure}
    \centering
\begin{lstlisting}[style=ACSLStyle,mathescape,escapebegin=\color{mGray}
]
#define LO 0
#define HI 10

extern int glob_stee_primary_status;  //input
extern int glob_stee_sndary_status;   //output

/*@
  ensures LO <= global_stee_primary_status <= HI &&      //$c_\mathit{havoc}$ 
  assigns global_stee_primary_status;                   
*/
void havoc_input() { ... }

/*@ ensures \result == glob_stee_status;                 //$c_\mathit{read}$ 
    assigns \nothing; */            
int rtdb_read_primary_stee_status() { ... } //$\beta_4$

/*@ ensures \old(stee_info) != 0 ==> \result == 0 &&     //$c_\mathit{eval}$ 
          \old(stee_info) == 0 ==> \result == 1;  
    assigns \nothing;  */
int evaluate_stee_status(int stee_info) { ... }

/*@ ensures glob_stee_primary_status == status;          //$c_\mathit{write}$
    assigns glob_stee_primary_status;  */      
void rtdb_write_sndary_stee_status(int status) { ... }

void steering() { //$\beta_3$
  //Read input data from the RTDB
  // 0 == Flawless; Non-zero == Error 
  int primary_info = rtdb_read_primary_stee_status();     //$s_\mathit{read}$
  
  //Evaluate the data
  int sndary_info = evaluate_stee_status(primary_info);   //$s_\mathit{eval}$
  
  //Write back the result of the evaluation to the RTDB
  rtdb_write_sndary_stee_status(sndary_info);             //$s_\mathit{write}$
}

//scheduler
void main() { //$\beta_1$
  while(1) {  //$\beta_2$
    havoc_input(); //simulating environment
    steering();
  }
}
\end{lstlisting}
    \caption{STEE example}
    \label{fig:stee_skeleton_code}
\end{figure}

\paragraph{Running example.}

We illustrate our approach on a simplified version of a software module  from the automotive industry. The module is called \Stee{}, and is intended to control the secondary steering of a heavy-vehicle, and, in case of malfunction of the primary steering module, activate the secondary steering functionality. The \Stee{} function is called periodically by the scheduler of the embedded system's control software. This example has been used in earlier work and is described in more detail in~\cite{ami-lid-gur-22-isola}. \autoref{fig:stee_skeleton_code} shows a code skeleton of \Stee{} and a main function which acts as the scheduler. The example is interesting in this context, as it combines the temporal behaviour of the scheduler (the while loop) with the transformational behaviour of \Stee{}. We use ACSL~\cite{acsl} to specify certain parts of the program with contracts. 

\paragraph{Related work.}
The closest work to ours is presented by Beyer et al.~\cite{beye-et-al-09-lbe}, which abstracts programs by automatically computing  \emph{summaries} of loop-free fragments of the code. The difference from our approach is that contracts may also specify code with loops and procedure calls. Also, the summaries are automatically extracted, whereas we assume contracts provided by, e.g., humans. A summary can, however, also be viewed as the strongest contract for the code block it is extracted from.
Beckert et al.~\cite{beckert-et-al-20-jml} have developed an approach for (bounded) model checking of Java programs with JML-contracts. The verification approach is similar to the one presented here, but model checking is carried out with a bounded model checker (JBMC), and they do not consider an intermediate (flow graph) representation.
The \textsc{Kratos2} tool~\cite{grig-jona-23-kratos} allows for converting C-programs into the intermediary K2 language, which can in turn be converted into \nuxmv{} models to be model checked. It does not, as far as we are aware, handle abstractions based on contracts or summaries. 

The \textsc{MoXI}~\cite{rozi-et-al-24-moxi} language is another intermediate language for symbolic model checking. \textsc{MoXI} is more general than the flow graphs presented here and is, for example, well-adapted also for hardware modelling. Flow graphs are intended specifically for modelling procedural software that has been deductively verified.


\paragraph{Outline.}

The paper is structured as follows.
In Section~\ref{sec:approach}, we present our overall approach to deductive verification based program abstraction. 
In Section~\ref{sec:progs-and-flow-graphs}, we describe our notation for programs (annotated with contracts) and define the notion of flow graph, which we use to represent program abstractions, while in
Section~\ref{sec:programs-to-flow-graphs} we show how to extract such flow graphs from annotated programs.
Once a flow graph is extracted, it can be model checked as described in Section~\ref{sec:model-checking-flow-graphs}. 
We discuss various aspects of our approach in
Section~\ref{sec:discussion}, and conclude with Section~\ref{sec:conclusion}.


\section{Deductive Verification Based Program Abstraction}
\label{sec:approach}

\definecolor{nodeGray}{rgb}{0.96,0.96,0.96}
\definecolor{cornerGray}{rgb}{0.90,0.90,0.87}
\definecolor{beaublue}{rgb}{0.74, 0.83, 0.9}
\definecolor{mistyrose}{rgb}{1.0, 0.89, 0.88}
\tikzstyle{nodeone} = [rounded corners, text width=1.7cm, minimum height=1.8cm,text centered, draw=black, fill = nodeGray, font = \scriptsize]
\tikzstyle{nodetwo} = [rounded corners, text width=1.5cm, minimum height=5.5cm, minimum width = 6.5cm, text centered, draw=black, fill = beaublue, font = \scriptsize]
\tikzstyle{cornernode} = [rectangle, below right,draw, fill=cornerGray]

\tikzstyle{arrow} = [thick,->,>=stealth]

In previous work~\cite{ami-lid-gur-22-isola}, we introduced a workflow and tool chain for software model checking of C programs, where we used contracts and deductive verification. Using contracts as a basis of abstraction, we could utilise the modularity of deductive verification in the otherwise monolithic model checking setting.

In that work, we relied on the TLA framework~\cite{lamp-94-tla} to act as the \emph{bridging agent} between the realms of deductive verification and model checking. Following this idea, we considered a translation of C programs, annotated with contracts, into $\tlaplus$~\cite{lamport-book-02} specifications. However, by depending heavily on TLA, we imposed certain limitations on the considered tool chain. For instance, the \tlc{} model checker, which is part of the TLA framework, is an explicit-state model checker, whereas one might expect a symbolic model checker to be more suitable for models related to \emph{symbolic transition systems}, such as the $\tlaplus$ specifications.

\begin{figure}[tb]
    \centering
\begin{tikzpicture}
\scriptsize
    \node[draw=black,rounded corners,align=center] (prog) {Program annotated \\ with contracts};
    \node[draw=black, rectangle, below = 1em of prog,align=left] (verify) {Verify contracts using\\ deductive verification};
    \node[draw=black, rounded corners, right = 1.5em of verify,align=left] (tofg) {Flow graph};
    \node[draw=black, rounded corners, above right = 0.7em and 1.5em of tofg,align=left,minimum width=2.05cm] (tla) {$\tlaplus$ model};
    \node[draw=black, rounded corners, right = 1.5em  of tofg,align=left,minimum width=2.05cm] (nuxmv) {\nuxmv{} model};
    \node[below right = 0.7em and 1.5em of tofg,align=left,minimum width=2.05cm] (other) {$\cdots\cdots\cdots$};
    \node[draw=black, rectangle, right = 1.5em of tla,align=left] (mc1) {Model check};
    \node[draw=black, rectangle, right = 1.5em of nuxmv,align=left] (mc2) {Model check};
    \draw[->] (prog) to (verify);
    \draw[->] (verify) to (tofg);
    \draw[->] (tofg) to (tla.south west);
    \draw[->] (tofg) to (nuxmv);
    \draw[->] (tofg) to (other.north west);
    \draw[->] (tla) to (mc1);
    \draw[->] (nuxmv) to (mc2);
\end{tikzpicture}
    
    \caption{Overview of the verification workflow using flow graphs}
    \label{fig:workflow_approach}
\end{figure}
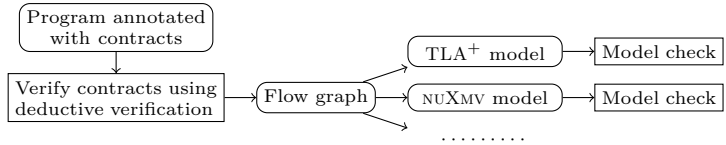

In this paper, we generalise our previous work by introducing an intermediate step, in which we first translate the program annotated with contracts, which we henceforth shall call \emph{annotated program}, into a representation as a \emph{flow graph}. The rationale for this is that this representation is independent of the source language, and can further be used to extract models in different modelling languages depending, for example, on which tool is most suitable for the verification task at hand. In this work, we consider extractions into $\tlaplus$ and the \nuxmv{} language, but we expect that also other modelling languages can be used. The verification approach, where flow graphs act as an intermediate step in the model extraction, is illustrated in \autoref{fig:workflow_approach}. In the workflow, the role of the deductive verification step is to ensure certain \emph{soundness} properties of the entire translation and verification chain.

Flow graphs are a formalism previously developed for the purposes of compositional model checking~\cite{sol-gur-16-scp}. We argue, for two main reasons, that they are also well-suited for the context presented here. First, there is a natural translation from a program annotated with contracts into a flow graph. Second, we believe that flow graphs have a natural translation into the modelling languages of several existing model checkers.





\section{Annotated Programs and Flow Graphs}
\label{sec:progs-and-flow-graphs}

In this section, we first present our definition of annotated programs as directed graphs (akin to the standard notion of \emph{control flow graphs}). We then define a similar object called a \emph{flow graph}, which allows certain aspects of the original directed graph of the program to be abstracted away.

\subsection{Annotated Programs}
\label{subsec:programs}

To simplify the presentation, we shall not consider here a specific programming language with its concrete syntax and semantics,
but shall instead assume that programs are given to us as directed graphs.
We consider programs as a set of procedures, which, in turn, consist of a set of (labelled) code blocks that map program locations to statements. With such a formalisation,  contracts may specify either the entire procedure or smaller blocks within a procedure. The idea is to have a representation of programs that captures both the data transformation aspects and control flow of a program that has been annotated with contracts. A program is thus seen as jumping between different control points, and at each control point some data transformation occurs (possibly the identity transformation). The data transformations stem either from program statements or from contracts.

The sets of statements~$\stmt$ and Boolean expressions~$\bexpr$ are left unspecified, and similarly the set of contracts~$C$. 

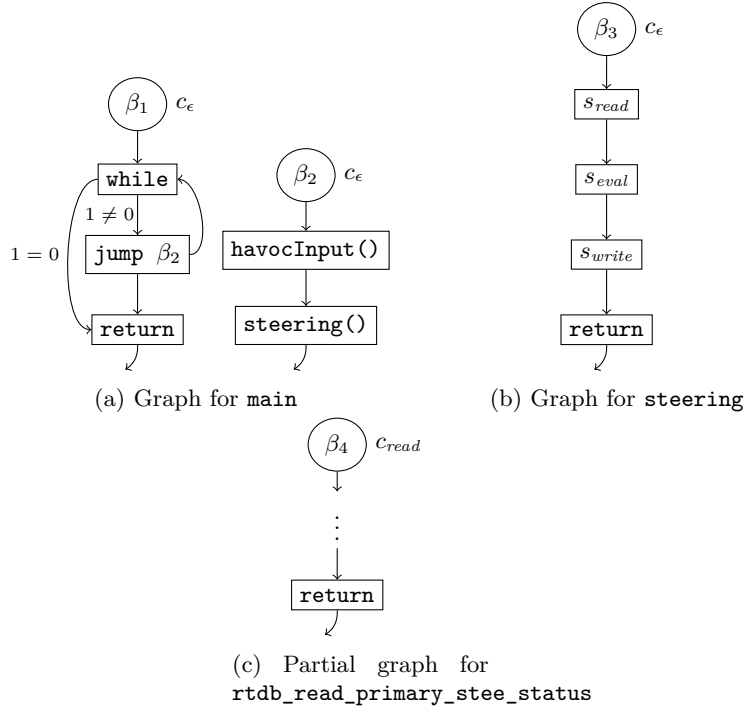
\begin{figure}[t]
    \centering
\begin{subfigure}{0.55\linewidth}
\centering
    \begin{tikzpicture}
        \node[pentrynode, label=right:$c_\epsilon$] (m0) {$\beta_1$};
        \node[pregnode,initial text = $\beta_1$,below of = m0] (m1) {\lstinline!while!};
        \node[pregnode,below of = m1] (m2) {\lstinline[mathescape]!jump $\beta_2$!};
        \pretnode{m6}{\lstinline[mathescape]!return!}{below of=m2}
        \draw[->] (m0) to (m1);
        \draw[->] (m1) to node[left,labelnode,align=left,xshift=2pt] {$1\neq 0$} (m2);
        \draw[->] (m2) to (m6);
        \draw[->,out=0, in=0,looseness=0.8,align=left] (m2) to (m1);
        \draw[->,out=180, in=180,looseness=0.6] (m1) to node[left,labelnode] {$1=0$} (m6);
    \end{tikzpicture}
    \begin{tikzpicture}
        \node[pentrynode, label=right:$c_\epsilon$] (m2) {$\beta_2$};
        \node[pregnode,below of = m2] (m3) {\lstinline!havocInput()!};
        \pretnode{st}{\lstinline!steering()!}{below of = m3}
        \draw[->] (m2) to (m3);
        \draw[->] (m3) to (st);
    \end{tikzpicture}
    \caption{Graph for \lstinline!main!}
\label{fig:method_graph_main}
\end{subfigure}
\begin{subfigure}{0.35\linewidth}
\centering
    \begin{tikzpicture}
        \node[pentrynode,label=right:$c_\epsilon$] (s0) {\lstinline[mathescape]!$\beta_3$!};
        \node[pregnode,below of = s0] (s1) {$s_\mathit{read}$};
        \node[pregnode={$a_{c_{eval}}$}, below of = s1] (s2) {$s_\mathit{eval}$};
        \node[pregnode={$a_{c_{write}}$}, below of = s2] (s3) {$s_\mathit{write}$};
        \pretnode{s4}{\lstinline!return!}{below of = s3};
        \draw[->] (s0) to (s1);
        \draw[->] (s1) to (s2);
        \draw[->] (s2) to (s3);
        \draw[->] (s3) to (s4);
    \end{tikzpicture}
    \caption{Graph for \lstinline!steering!}
\label{fig:mathod_graph_stee}
\end{subfigure}
\begin{subfigure}{0.28\linewidth}
    \centering
    \begin{tikzpicture}
        \node[pentrynode, label=right:{$c_\mathit{read}$}] (s0) {\lstinline[mathescape]!$\beta_4$!};
        \node[below of = s0] (s1) {$\vdots$};
        \pretnode{s2}{\lstinline[mathescape]!return!}{below of = s1};
        \draw[->] (s0) to (s1);
        \draw[->] (s1) to (s2);
    \end{tikzpicture}
    \caption{Partial graph for \lstinline!rtdb_read_primary_stee_status!}
\label{fig:mathod_graph_stee}
\end{subfigure}
    \caption{Directed graph representation of parts of the program in \autoref{fig:stee_skeleton_code}}
    \label{fig:stee_graph}
\end{figure}

\begin{definition}[Annotated Code Block]
    An annotated code block is a tuple $\beta = (V_\beta, E_\beta, \pi_{V_\beta}, \pi_{E_\beta}, v_{e_\beta}, v_{r_\beta}, C_\beta)$, where 
$V_\beta$ is a set of control points, 
$E_\beta \subseteq V_\beta \times V_\beta$ a set of edges, 
$\pi_{V_\beta}: V_\beta \rightarrow \stmt$  a labelling of control points with statements, and $\pi_{E_\beta}: E_\beta \rightarrow \bexpr$ a labelling of edges with path conditions. Finally, $v_{e_\beta} \in V_\beta$ is the entry point and $v_{r_\beta} \in V_\beta$ the exit point of the block, and $C_\beta$ is the contract for the block.
\end{definition}

In the set of contracts, we let $c_\epsilon$ denote the empty contract, indicating that the respective block has \emph{not} been specified. 
%
The relation~$E_\beta$ describes the flow of control between statements, and the labelling $\pi_{E_\beta}$ is induced from the Boolean guards of conditional or loop statements. 
In the set of Boolean expressions $\bexpr{}$, we include an empty expression $b_\epsilon$. When $\pi_{E_\beta} (v, v') = b$ and $b \neq b_\epsilon$, we say that $v'$ is \emph{guarded} by~$b$ (or that $b$ is a \emph{guard}). The $\mathsf{jump}$ statement corresponds to (intra-procedural) conditional branching, such as if-then-else and while statements.


In the example program from \autoref{fig:stee_skeleton_code}, we consider the contracts to annotate the entry block statement of the respective procedure body. A visual depiction of parts of the graph representation of the program is shown in \autoref{fig:stee_graph}. 

\begin{definition}[Annotated Procedure]
An \emph{annotated procedure} with name~$p$ is a pair $P_p = (\mathcal{B}_p, \beta_{p})$, where $\mathcal{B}_p$ is a collection of annotated blocks, and $\beta_{p}$ is the entry block.
%
\end{definition}


\begin{definition}[Annotated Program]
An \emph{annotated program} consists of a pair $(\mathcal{P}, \mathit{main})$, where $\mathcal{P}$ is a collection of annotated procedures, and $\mathit{main}$ the name of the starting procedure. 
\end{definition}
%
%
%
The definition of an annotated program puts no constraint on the correctness of the program relative to its annotations (that is, whether every annotated part implements its companion contract or not). 
However, as described in \autoref{sec:approach}, verifying that the code fulfils the contracts is key to ensuring soundness of the abstraction approach considered in this paper. Thus, we shall say that a program is \emph{correctly annotated} if every annotated block fulfils the contract annotation (which can be verified using, e.g., deductive verification tools).



\subsection{Flow Graphs}
\label{sec:flow-graphs}

We proceed with a formal definition of flow graphs, adapted, with significant modifications, from~\cite{sol-gur-16-scp}.

First, we define the set of program states as $\state = \state^L \times \state^G$, where the local states $\state^L$ and global states $\state^G$ capture the values of the local and global variables, respectively. Given a state $s \in \state$, we denote with $\sloc$ its component in $\state^L$ and with $\sglob$ its component in $\state^G$.
%
Next, we define the notion of an \emph{action}. We take the view of the TLA framework and evaluate actions on pairs of states, thus viewing them as relations between pre-states and post-states (similar to the well-known notion of a state transformer).

\begin{definition}[Action]
\label{def:action}
An \emph{action} $a \in A$ is a Boolean expression over \emph{primed} variables~$x'$ and \emph{non-primed} variables~$x$.
\end{definition}
We do not provide the syntax in full detail here, but follow the idea of the TLA framework, where actions are logic formulas with variables being either \emph{primed} or \emph{non-primed}, with the intention that non-primed variables are evaluated in the pre-state and primed variables in the post-state. For example, $x' = x+1$ is an action that holds over all state pairs $(s,s')$ such that $s'(x) = s(x) + 1$. Semantically, an action is defined as a binary relation on states: $\denot{a}\subseteq \state \times \state$.

\begin{definition}[Procedure Flow Graph]
\label{def:proc-graph}
A \emph{procedure flow graph}, for a procedure with name~$p$, is a tuple $F_p = (N_p, D, \rightarrow_p, A, \lambda_p, n_{p_e}, n_{p_r})$, where $N_p$ is a finite set of nodes, $D$ a set of procedure names, $\rightarrow_p \subseteq N_p \times (D \cup \set{\epsilon}) \times N_p$ a labelled transition relation, $A$ a set of actions, $\lambda_p: N_p \rightarrow A$ a labelling of nodes with actions, $n_{p_e} \in N_p$ the entry node, and $n_{p_r} \in N_p$ the return node. 
\end{definition}

By labelling nodes with actions, each node can be seen as a state transformer, induced by either program statements or block contracts. Edges are either \emph{call transitions}, labelled with a procedure name, or \emph{silent transitions}, labelled with~$\epsilon$. Call transitions correspond to to procedure calls and silent transitions to intra-procedural transfer of control.

\begin{definition}[Flow Graph]
\label{def:flow-graph}
    A flow graph is a pair $(\mathcal{F}, \mathit{main})$, where $\mathcal{F}$ is a collection of procedure flow graphs, and ${\mathit{main}}$ is the name of the main procedure flow graph.
\end{definition}
An example of a flow graph is shown in \autoref{fig:stee_skeleton_flow_graph}, which is described in detail in \autoref{sec:programs-to-flow-graphs}.



We now proceed by defining the operational semantics of flow graphs by means of \emph{pushdown systems}. We first introduce pushdown systems formally.

\begin{definition}[Pushdown System]
    A \emph{pushdown system (PDS)} is a tuple $\mathcal{S} = (S, \Gamma, \triangle, I)$, where $S$ is a set of control states, $\Gamma$ a stack alphabet, and $\triangle \subseteq (S \times \Gamma) \times (S \times \Gamma^*)$ a set of rewrite rules. Moreover, we refer to $S \times \Gamma^\ast$ as the set of \emph{configurations}, and $I \subseteq S \times \Gamma^\ast$ is the set of initial configurations.
\end{definition}
Let $\langle s,\gamma \rangle \rightarrow \langle s', \omega \rangle \in \triangle$ be a rewrite rule. Then, for each $\omega' \in \Gamma^*$, $\langle s', \omega \cdot \omega' \rangle$ is called an \emph{immediate successor} of $\langle s, \gamma\cdot\omega' \rangle$.

\begin{definition}[Run of a PDS]
    A \emph{run} of a pushdown system $\mathcal{S}$ is an infinite sequence of configurations $\langle s_1, \omega_1\rangle \langle s_2, \omega_2 \rangle\ldots$ where $\langle s_1, \omega_1\rangle \in I$, and for every $i \geq 1$, $\langle s_{i+1}, \omega_{i+1} \rangle$ is an immediate successor of $\langle s_i, \omega_i \rangle$.
    Moreover, a \emph{state-run} of~$\mathcal{S}$ is an infinite sequence of states $s_1s_2\ldots$ such that there exists a run of~$\mathcal{S}$ of the form  $\langle s_1, \omega_1 \rangle \langle s_2, \omega_2 \rangle\ldots$ for some $\omega_1, \omega_2, \ldots \in \Gamma^\ast$.
\end{definition}




We are now ready to define how a flow graph \emph{induces} a pushdown system, as inspired by Schwoon~\cite{schwo-02-phd}, so that runs of the flow graph can be defined in terms of state-runs of the induced pushdown system.

\begin{definition}[Induced PDS]
\label{def:induced-pds}
Given a set of initial global states $G_\mathit{init} \subseteq \state^G$, a flow graph~$(\mathcal{F}, \mathit{main})$ \emph{induces} a pushdown system $\mathcal{S}_\mathcal{F} = (S, \Gamma, \triangle, I)$, where:
\begin{itemize}[label=$\bullet$]
    \item $S = \state^G$, i.e., the global states of the program serve as control states;
    \item $\Gamma = (\bigcup_{F_p \in \mathcal{F}}N_p) \times \state^L$, i.e., we store on the stack pairs of nodes and local states of procedures;
    \item $\Delta$ is induced as follows. Every procedure flow graph $F_p$ induces the following sets of rewrite rules, where $\locinit(p) \in \state^L$ captures the initial values of the local variables in $p$:
    \begin{enumerate}[label=(\roman*)]
        \item For each silent transition $(n_{p_1}, \epsilon, n_{p_2})$, the set of rewrite rules:
        \begin{align*}
        \{\langle \glob{s}, (n_{p_1},\sloc) \rangle \rightarrow \langle \glob{t}, (n_{p_2}, \loc{t})  \rangle \mid 
        (s, t) \in \denot{\lambda_p(n_{p_1})} \}
                \end{align*}
        \item For each call transition $(n_{p_1}, q, n_{p_2})$, the set of rewrite rules:
        \begin{align*}
        \{\langle \glob{s}, (n_{p_1}, \loc{s}) \rangle \rightarrow \langle \glob{t}, (q_\mathit{e}, \locinit(q)) \cdot (n_{p_2}, \loc{t}) \rangle \mid 
         (s,t) \in \denot{\lambda_p(n_{p_1})}\}
        \end{align*}
        \item If $p \neq \mathit{main}$, then for the return node $n_{p_r}$, the set of rewrite rules:
         \begin{align*}
         \{\langle s^g, (n_{p_r}, l) \rangle \rightarrow \langle t^g, \epsilon \rangle \mid (s, t) \in \denot{\lambda_p(n_{p_r})} \}
        \end{align*}
    \end{enumerate}
The set of transition rules $\Delta$ is then the union of the above rules for each transition node and the return node. 
    \item $I = \{ \langle s^g, (\mathit{main}_e, \locinit(\mathit{main}))\rangle \mid s^g \in G_{\mathit{init}}\}$.
\end{itemize}
\end{definition}

The overall idea of the induced pushdown system is that we execute (the action of) the current node, and then make a transition to a possible next edge. For return nodes, the transition will always be back to the call site of the caller function. For non-return nodes, a silent transition means a transition directly to the target, and a call transition means that we transition to the entry of the function being called.
The control flow for procedure calls and returns is modelled using a stack. For a call transition $(n_{p_1}, q, n_{p_2})$ in procedure $p$, we push the next node in the current procedure graph (i.e., $n_{p_2}$) and the current local variable state to the stack. Then, for the case when the procedure $q$ returns, we simply add a rule that pops the local state and node from the stack (and discards the current local state), so that the execution of $p$ can resume from $n_{p_2}$ with correct values for the local variables. For silent transitions, we simply let the stack be untouched. The initial configurations are given by the possible initial global states, together with the entry node of the main procedure.

\begin{definition}[Run of a Flow Graph]
A \emph{run} of a flow graph~$(\mathcal{F}, \mathit{main})$ is a state-run of the pushdown system induced by~$\mathcal{F}$. 
\end{definition}
That is, when defining the run of the flow graph, we consider only the global component of the state.
We view linear-time properties of a flow graph~$\mathcal{F}$ as properties of its runs. Observe that we only consider infinite sequences of the induced push-down system, which implicitly imposes the constraint that flow graphs are total. That is, each node in the flow graph has at least one outgoing edge. As we shall show in the next section, 

To illustrate, consider the flow graph in \autoref{fig:stee_skeleton_flow_graph}. 
The following sequence of configurations shows the first three steps of a run in the induced pushdown system. 
\begin{align*}
    \langle s_{1}^g, (n_{m_1},s_{1}^l) \rangle\, \langle s_{2}^g, (n_{m_2},s_{2}^l) \rangle\, \langle s_{2}^g, 
    (n_{s_1},\mathit{init}^l(\mathit{stee})) \cdot
    (n_{m_3},s_{2}^l) \rangle \ldots
\end{align*}
where $s_{1}^l = \mathit{init}^l(\mathit{main})$, $s_{1}^g$ assigns some initial values to the global variables, and $(s_1, s_2) \in \denot{a_{c_\mathit{havoc}}}$. The corresponding induced state-run (and thereby run of the flow graph) is the sequence $s_{1}^g s_{2}^g s_{2}^g \ldots$.


\subsection{From Annotated Programs to Flow Graphs}
\label{sec:programs-to-flow-graphs}
In this section, we sketch a general translation from annotated programs into flow graphs, which was already hinted at in \autoref{sec:progs-and-flow-graphs}, where we showed in \autoref{fig:stee_skeleton_flow_graph} the flow graph representation of \Stee{}. We leave a fully formalised definition of the translation as future work. 


\begin{figure}
    \centering
\begin{subfigure}{0.45\linewidth}
\centering
    \begin{tikzpicture}
    \node[entrynode={$a_{c_{\mathit{havoc}}}$},text=white] (m1) {{$n_{m_1}$}};
        \node[regnode={$\idaction\, \wedge$ \\ $1 \neq 0$},below of=m1,text=white] (m2) {{$n_{m_2}$}};
        \node[regnode={$\idaction$},below of=m2,text=white] (m3) {{$n_{m_3}$}};
        \node[retnode={\\[8pt] $\idaction \wedge$\\$ 1 = 0$},below of=m3,text=white] (m4) {{$n_{m_4}$}};
        \draw[->] (m1) to node[left,labelnode] {$\epsilon$} (m2);
        \draw[->] (m2) to node[left,labelnode,align=left,xshift=2pt] {\lstinline!stee-!\\\lstinline!ring!} (m3);
        \draw[->,out=180, in=180,looseness=1.4,align=left] (m3) to node[left,labelnode] {$\epsilon$} (m1);
        \draw[->,out=330, in=30,looseness=1.7] (m1) to node[right,labelnode] {$\epsilon$} (m4);
        \draw[->,loop below] (m4) to (m4);
    \end{tikzpicture}
    \caption{Flow graph for \lstinline!main!}
\label{fig:method_graph_main}
\end{subfigure}
\begin{subfigure}{0.45\linewidth}
\centering
    \begin{tikzpicture}
        \node[entrynode={$a_{c_{read}}$},text=gray!20] (s1) {{$n_{s_1}$}};
        \node[regnode={$a_{c_{eval}}$}, below of = s1,text=white] (s2) {{$n_{s_2}$}};
        \node[regnode={$a_{c_{write}}$}, below of = s2,text=white] (s3) {{$n_{s_3}$}};
        \node[retnode={$\idaction$}, below of = s3,text=white] (s4) {{$n_{s_4}$}};
        \draw[->] (s1) to node[right,labelnode] {$\epsilon$} (s2);
        \draw[->] (s2) to node[right,labelnode] {$\epsilon$} (s3);
        \draw[->] (s3) to node[right,labelnode] {$\epsilon$} (s4);
    \end{tikzpicture}
    \caption{Flow graph for \lstinline!steering!}
\label{fig:mathod_graph_stee}
\end{subfigure}
\caption{Flow graph for \Stee{}}
\label{fig:stee_skeleton_flow_graph}
\end{figure}

For the translation sketch, we assume given a canonical representation of statements, contracts and Boolean expressions as actions, equivalent to their standard denotational semantics, and use $a_c$, $a_s$ and $a_b$ to denote the actions corresponding to the contract~$c$, statement~$s$, and Boolean guard~$b$, respectively. 
Then, an annotated program $(\mathcal{P}, \mathit{main})$ is translated into a flow graph as follows. 

For each annotated procedure $P_p = (\mathcal{B}_p, \beta_{p})$, 
in the program, we construct a procedure flow graph $F_p = (N_p, D, \rightarrow_p, A, \lambda_p, n_{p_e}, n_{p_r})$. The nodes~$N_p$ are initially given by the control points, i.e., $\bigcup_{\beta \in \mathcal{B}}V_\beta$.
The labelling $\lambda_{p}$ of a node~$n_v$ corresponding to a control point~$v$ is defined as follows:
\begin{align*}
\lambda_p(n_v) = \begin{cases}
   a_{c_\beta} & \text{if } \pi_V(v) = (\mathtt{jump}\;\beta) \text{ and } c_\beta \neq c_\epsilon ;  \\
   a_{c_\beta} & \text{if } \pi_V(v) = (s) \text{ and $s$ is a procedure call to $p$ where the entry}
   \\
   & \text{\hspace*{5.8em}block $\beta_p$ is annotated with contract $c_{\beta_p} \neq c_\epsilon$} ;
   \\
   a_s & \text{if } \pi_V(v) = (s) \text{ and } s \text{ is an assignment statement} ; \\
   \idaction & \text{otherwise} .
\end{cases}
\end{align*}
That is, we replace annotated blocks, and calls to annotated procedures, with  their respective contract. 
In addition to the labelling defined above, we also merge any remaining jumps, i.e., replace jumps to a non-annotated block $\beta$ with the body of $\beta$. 
Lastly, if $s$ is guarded by~$b$, then~$a_b$ is added as a conjunct to the label. 

The sources and targets of the transition relation $\rightarrow_p$ for the remaining nodes are defined canonically from~$E_\beta$ for each block $\beta$. The labelling~$e$ of an edge  $(n_{v_i}, e, n_{v_j})$ is defined as follows:
\begin{align*}
    e = \begin{cases}
        p & \text{if } \pi_V(v_i) = (s) \text{ and $s$ is a call to procedure $p$ where the entry}\\ 
        & \text{block $\beta_{p}$ of $p$ is annotated with $c_\epsilon$} ;
        \\
        \epsilon & \text{otherwise} .
    \end{cases}
\end{align*}
That is, we keep as edge labels all calls to procedures where the body is not specified. Lastly, we add an $\epsilon$-edge from the return node of the $\mathit{main}$ procedure to itself, thus modelling terminating executions as ending by stuttering indefinitely in the return node of $\mathit{main}$.



In the example in \autoref{fig:stee_skeleton_flow_graph}, we see that the nodes $n_{m_1}$, $n_{s_1}$, $n_{s_2}$, and $n_{s_3}$ are labelled with (the action of) the respective contract. We assume here for simplicity that the actions $a_{c_\mathit{read}}$ and $a_{c_\mathit{eval}}$ also describe the assignments to the local variables \lstinline$primary_info$ and \lstinline$sndary_info$. The remaining nodes are labelled with the identity action, and the guarded statements stemming from the while loop in the main procedure have a conjunct for the guards. The only non-specified procedure call, the one to \lstinline!steering!, is maintained as the label of the transition from $n_{m_2}$ to $n_{m_3}$.

\section{Model Checking of Flow Graphs}
\label{sec:model-checking-flow-graphs}

Once a flow graph has been extracted from an annotated program, it can be model checked against temporal properties by translating it into the input modelling language of the specific model checking framework at hand. 
Here, we illustrate the idea using two well-known frameworks, TLA and \nuxmv{}. 

\subsection{Model Checking with \tlc{}}
\tlc{}~\cite{yu-et-al-99-tlc} is an explicit state model checker that is part of the TLA~\cite{lamp-94-tla} framework. \tlc{} takes as input a $\tlaplus$~\cite{lamport-book-02} program, where $\tlaplus$ is the language implementation of TLA. \tlc{} can model check it either for deadlocks, or against temporal properties, which are also specified in $\tlaplus$. A beautiful aspect of \tlc{} is that both the models and the properties are specified in the same language. The $\tlaplus$ language uses actions (see \autoref{sec:flow-graphs}) as elementary propositions, and temporal properties are specified as in LTL, using, e.g., the temporal operators $\lozenge$ (``\emph{eventually}'') and $\square$ (``\emph{always}''). Programs in $\tlaplus$ are defined as \emph{symbolic transition systems}, comprised of a unary \emph{initial-state} predicate $\mathit{Init}$ and a binary \emph{next-state} predicate~$M$. Sometimes, a third component, a \emph{fairness constraint} $F$ is added explicitly as well. Given a triple $(\mathit{Init}, M, F)$, the corresponding $\tlaplus$ program is the formula $\mathit{Init} \wedge \square M \wedge F$.


To model check a flow graph with \tlc{}, we translate the flow graph to a semantically equivalent $\tlaplus$ specification. In particular, we define a variable~$n$ representing the current node in a run, and a variable $\mathit{st}$ representing the stack in the pushdown system. We also define $s^g$ and $s^l$ corresponding to the state for the local and global variables. The stack is modified with macros $\mathit{push}$ and $\mathit{pop}$, where $\mathit{push}(x, \mathit{st})$ returns the stack obtained by pushing $x$ to $\mathit{st}$, and $\mathit{pop}(\mathit{st})$ returns a pair $(x, \mathit{st}')$, where $x$ is the top element of $\mathit{st}$ and $\mathit{st}'$ is the obtained by popping $x$ from $\mathit{st}$.

\begin{definition}[Induced $\tlaplus$ Specification]
\label{def:induced-tlaplus-spec}
Given a flow graph $(\mathcal{F}, \mathit{main})$, we define, for each procedure flow graph~$F_p \in \mathcal{F}$, a set of actions as follows:
\begin{enumerate}[label=(\roman*)]
    \item For each silent transition $(n_{p_1}, \epsilon, n_{p_2})$, an action:
    $$ n = n_{p_1} \wedge n' = n_{p_2} \wedge st' = st \wedge \lambda_p(n_{p_1})$$
    \item For each call transition $(n_{p_1}, p', n_{p_2})$, where $p'$ is a procedure with entry node $n_{p_e'}$, an action:
     $$ n = n_{p_1} \wedge n' = n_{p_e'} \wedge st' = \mathit{push}((n_{p_2}, s^l), st) \wedge \lambda_p(n_{p_1})
     $$
     \item For the return node $n_{p_r}$, an action:
     $$ ((n', (s^l)'), \mathit{st}') = \mathit{pop}(\mathit{st}) \wedge \lambda_p(n_{p_r}) $$
\end{enumerate}
The full $\tlaplus$ specification of the program is then the specification:
$$\mathit{Init} \wedge \square M$$ 
where $M$ is the disjunction of all actions induced as described above, and $\mathit{Init}$ is the initial constraint on the global variables that sets~$n$ and the local variables according to $F_\mathit{main}$.
\end{definition}
\begin{figure}[htb!]
    \centering
\begin{lstlisting}[language=TLA]
\* Contracts
c_havoc == glob_stee_primary_status' \in LO..HI /\
           UNCHANGED(glob_stee_sndary_status)
c_read(res_var) == assign_loc(res_var, glob_stee_primary_status)
                   /\ fix_globs
c_eval(stee_info, res_var) ==
  (fetch_loc(stee_info) # 0 => assign_loc(res_var, 0))
  /\ (fetch_loc(stee_info) = 0 => assign_loc(res_var, 1))
  /\ fix_globs
c_write(status) == (glob_stee_sndary_status' = fetch_loc(status))
                   /\ fix_lvars /\ UNCHANGED(glob_stee_primary_status)

\*Main
main1 == n = "n_main1" /\ n' = "n_main2" /\ c_havoc /\ fix_stack /\ fix_lvars
main2 == n = "n_main2" /\ n' = "n_main4" /\ fix_stack /\ fix_lvars /\ 1 = 0
main3 == n = "n_main2" /\ n' = "n_stee1"  /\
         push("n_main3", lvars) /\ init_stee(lvars') /\ fix_globs /\ 1 # 0
main4 == n = "n_main3" /\ n' = "n_main1" /\ fix_stack /\ fix_lvars
         /\ fix_globs
main5 == n = "n_main4" /\ n' = "n_main4" /\ fix_stack /\ fix_lvars
         /\ fix_globs

\*Stee
stee1 == n = "n_stee1" /\ n' = "n_stee2" /\ fix_stack /\ c_read(primary_info)
stee2 == n = "n_stee2" /\ n' = "n_stee3" /\ fix_stack
         /\ c_eval(primary_info, sndary_info)
stee3 == n = "n_stee3" /\ n' = "n_stee4" /\ fix_stack /\ c_write(sndary_info)
stee4 == n = "n_stee4" /\ pop(n', lvars') /\ fix_globs

\*Program
Init == n = "n_main1" /\ init_main(lvars) /\ init_stack /\ init_globals
M == (main1 \/ main2 \/ main3 \/ main4 \/ main5 \/
      stee1 \/ stee2 \/ stee3 \/ stee4)

Spec == Init /\ [][M]_all_vars
\end{lstlisting}
    \caption{$\tlaplus$ model for \Stee{}}
    \label{fig:tlaplus_spec_stee}
\end{figure}
In \autoref{fig:tlaplus_spec_stee}, we show the key parts of the $\tlaplus$ model induced by the flow graph in \autoref{fig:stee_skeleton_flow_graph}. The full code is available in Appendix~\ref{app:full_models}.

As \tlc{} is an explicit-state model checker, model checking of the program is only possible for finite domains for the variables and the stack. In particular, we cannot model check flow graphs with infinite variable domains or unbounded recursion.

\subsection{Model Checking with \nuxmv{}}

\nuxmv~\cite{cava-et-al-14-xmv} is a symbolic model checker for finite or infinite-state systems, which takes as input a model in the \nuxmv{} language, and can verify the model against properties written in either LTL or CTL. Similar to $\tlaplus$, a \nuxmv{} model can also be given as a symbolic transition system, with an \emph{initial constraint} and \emph{transition constraints}, the latter corresponding to the next-state relation in $\tlaplus$. Transition constraints are, semantically speaking, binary relations over the states of the transitions system, and are thus similar to $\tlaplus$ actions. Syntactically, they differ from $\tlaplus$ actions in that they use the \lstinline!next! keyword instead of primed variables. For readability, \nuxmv{} also allows \emph{define} declarations, which function as macros.

The language 
On the other hand, \nuxmv{} supports model checking of infinite-state models, so that we may consider infinite domains for the program states.  

\begin{figure}[t]
    \centering
\begin{lstlisting}[language=SMV]
-- contracts
  c_havoc := next(glob_stee_primary_status) in 0..10
             & next(glob_stee_sndary_status) = glob_stee_sndary_status;
  c_read := next(lvars[primary_info]) = glob_stee_primary_status
            & fix_globs;
  c_eval := (lvars[primary_info] != 0 -> next(lvars[sndary_info]) = 0)
            & (lvars[primary_info] = 0 -> next(lvars[sndary_info]) = 1)
            & fix_globs;
  c_write := (next(glob_stee_sndary_status) = lvars[sndary_info])
             & fix_lvars
             & next(glob_stee_primary_status) = glob_stee_primary_status;


-- main
  Main1 := (n = n_main1 & next(n) = n_main2 & c_havoc & fix_lvars 
            & fix_stack);
  Main2 := (n = n_main2 & next(n) = n_main4 & fix_lvars & fix_stack & 1 = 0);
  Main3 := n = n_main2 & next(n) = n_main3 & fix_stack & fix_lvars & push
           & init_stee_lvars & 1 = 0;
  Main4 := n = n_main3 & next(n) = n_stee1 & fix_stack & fix_lvars & fix_globs;
  Main5 := n = n_main4 & next(n) = n_main4 & fix_stack & fix_lvars & fix_globs;

-- stee
  Stee1 := n = n_stee1 & next(n) = n_stee2 & fix_stack & c_read;
  Stee2 := n = n_stee2 & next(n) = n_stee3 & fix_stack & c_eval;
  Stee3 := n = n_stee3 & next(n) = n_stee4 & fix_stack & c_write;
  Stee4 := n = n_stee4 & pop & fix_globs;

INIT
  lvars = init_main_lvars & vstack = init_vstack & curr_top = 0 &
  nstack = init_nstack & init_globals & n = n_main1;

TRANS
  Main1 | Main2 | Main3 | Main4 | Main5 | Stee1 | Stee2 | Stee3 | Stee4;
\end{lstlisting}
    \caption{\nuxmv{} model of \Stee{}}
    \label{fig:stee_nuxmv_model}
\end{figure}

The following definition shows how a flow graph induces a \nuxmv{} model. Since both $\tlaplus$ and \nuxmv{} are symbolic transition systems, the translation below closely resembles \autoref{def:induced-tlaplus-spec}.

\begin{definition}[Induced \nuxmv{} Model]
\label{def:induced-nuxmv-model}
Given a flow graph $(\mathcal{F}, \mathit{main})$, we define, for each procedure flow graph~$F_p \in \mathcal{F}$, declarations as follows:

\begin{enumerate}[label=(\roman*)]
    \item For each silent transition $(n_{p_1}, \epsilon, n_{p_2})$, a define declaration:
    $$n = n_{p_1} \wedge \nunext{n} = n_{p_2} \wedge \nunext{\mathit{st}} = \mathit{st} \wedge \lambda_p(n_{p_1})$$
    \item For each call transition $(n_{p_1}, p', n_{p_2})$, where $p'$ is a procedure with entry node $n_{p_e'}$, an action:
     $$n = n_{p_1} \wedge \nunext{n} = n_{p_e'} \wedge \nunext{st} = \mathit{push}((n_{p_2}, s^l), st) \wedge \lambda_p(n_{p_1})
     $$
     \item For the return node $n_{p_r}$, a define declaration: 
     $$\nunext{((n, s^l), \mathit{st})} = \mathit{pop}(\mathit{st}) \wedge \lambda_p(n_{p_r}) $$
\end{enumerate}
The \nuxmv{} model is then given as a transition constraint corresponding to the disjunction of the define declarations in (i) - (iii), and an \lstinline!INIT! constraint, which describes the initial constraints of the global variables, and sets the initial node and local variables according to $F_\mathit{main}$.
\end{definition}

The key parts of the \nuxmv{} model for the flow graph in \autoref{fig:stee_skeleton_flow_graph} are shown in \autoref{fig:stee_nuxmv_model}. The full model is shown in Appendix~\ref{app:full_models}.



\section{Discussion}
\label{sec:discussion}

%
We expect the approach presented in this paper to be particularly applicable to automotive embedded software. To see why, consider the simplified \Stee{} example in \autoref{fig:stee_skeleton_code}. In a real example, the scheduler would periodically call numerous modules. Thus, if we annotate each module with a contract, we can perform model checking over the contracts of the modules, taking into account the actual code only of the scheduler.
This also aligns, in our experience, with how requirements are specified in the automotive industry, where software modules are typically accompanied with a number of safety-critical requirements that can be deductively verified. 

When considering the \Stee{} example, as presented here, the use of a pushdown system in the formalisation may seem unnecessary. However, depending on the property to verify, one may want to abstract away, for example, only a subset of the procedures in the program model. Then, the pushdown system is necessary to capture the behaviour of procedure calls which one does not want to abstract away from.

In the translations considered in this work, the role of 
deductive verification is to justify the abstraction for the model checking task. That is, by showing that the code satisfies the contracts, we may guarantee certain soundness properties of the abstraction. We are yet to work out the details of such guarantees, in particular the interplay between deductive verification and the temporal properties to be verified with model checking. 

In this work, we assumed that the contracts have been provided by some (likely human) oracle. By taking this approach, we allow the \emph{reuse} of earlier results in one domain (deductive verification), for the purpose of abstraction in another domain (model checking). The validity of this approach is corroborated by experience from an earlier case study on deductive verification of automotive embedded software~\cite{ung-et-al-24-stee}, where we encountered requirements that we formalised and verified using deductive verification, and other requirements that were better suited for verification with model checking.


\section{Conclusion}
\label{sec:conclusion}
In this paper, we described an approach for deductive verification based program abstraction, where \emph{flow graphs} act as an intermediate representation in extracting models from programs annotated with contracts. We have illustrated the approach on a code skeleton inspired by code from the heavy-vehicle industry, showing both how a program, annotated with contracts, can be translated into an abstract flow graph, and how the flow graph, in turn, induces $\tlaplus$ and \nuxmv{} models. We have also presented an operational semantics for flow graphs by means of an induced pushdown system, which allows semantic preservation properties of the translations to be established formally.

\paragraph{Future work.}
Future work includes first and foremost experimental evaluation of model extraction. In particular, we are interested in exploring how the abstraction impacts verification times when model checking. We also plan to show semantic preservation properties of the defined translations, and identify fragments of LTL for which the verification (model checking) in our approach is sound. Future work also includes implementing the translation steps described. Lastly, we also consider combining the contracts-based method presented here with the large-block encoding presented in \cite{beye-et-al-09-lbe}, to also consider abstraction based on \emph{summaries}, which can be automatically extracted from the code.

\paragraph{Acknowledgement.}

Some of the concepts presented here were discussed at the Lorentz center workshop on Contract Languages, 4-8 March 2024.


%
%
%
\bibliographystyle{splncs04}
\bibliography{references}

\newpage
\appendix

\section{Full $\tlaplus$ and \nuxmv{} model of \Stee{}}
\label{app:full_models}

\subsection{$\tlaplus$ model}
In the $\tlaplus$ model, the stack consists of a tuple of stacks, one that maintains the node and one that maintains the variables. The model also defines a number of auxiliary actions, including popping from and pushing to the stack, and reading from and writing to variables. The key parts of the model are the contract actions, which essentially capture the contracts as state transformers, and the program, which is based on applying \autoref{def:induced-tlaplus-spec} to \autoref{fig:stee_skeleton_flow_graph}. Note that each contract action and program action must define also what variables are \emph{not} written to, as required by TLC.

\begin{lstlisting}[language=TLA]
-----------------MODULE Stee_isola24------------------------------
EXTENDS Sequences, TLC, Naturals

VARIABLES
    nstack, vstack,
    \* Global vars
    glob_stee_primary_status,
    glob_stee_sndary_status,
    \* Current control state and local vars
    n, lvars

\*Constants
LO == 0
HI == 10
stack == <<nstack, vstack>>
primary_info == 1
sndary_info == 2

\*Inits
init_main(l) == l = <<>>
init_stee(l) == l = primary_info :> 0 @@ sndary_info :> 0

init_globals == glob_stee_primary_status \in {0,1} /\
                glob_stee_sndary_status \in {0,1}
init_stack == nstack = <<>> /\ vstack = <<>>

\*Auxiliary funcs
pop(_n, _v) == _n = Head(nstack) /\ _v = Head(vstack) /\ nstack' = Tail(nstack) /\ vstack' = Tail(vstack)
push(_n, _v) ==  nstack' = <<_n>> \o nstack /\ vstack' = <<_v>> \o vstack
assign_loc(x, v) == lvars' = [lvars EXCEPT ![x] = v] \*loc_env' = [loc_env EXCEPT !.lvars = [loc_env.lvars EXCEPT ![x] = v]]
fetch_loc(x) == lvars[x]
fix_globs == UNCHANGED(<<glob_stee_primary_status, glob_stee_sndary_status>>)
fix_stack == UNCHANGED(stack)
fix_lvars == UNCHANGED(lvars)
all_vars == <<glob_stee_primary_status, glob_stee_sndary_status, stack,lvars, n>>


\* Contracts
c_havoc == glob_stee_primary_status' \in LO..HI /\
           UNCHANGED(glob_stee_sndary_status)
c_read(res_var) == assign_loc(res_var, glob_stee_primary_status)
                   /\ fix_globs
c_eval(stee_info, res_var) ==
  (fetch_loc(stee_info) # 0 => assign_loc(res_var, 0))
  /\ (fetch_loc(stee_info) = 0 => assign_loc(res_var, 1))
  /\ fix_globs
c_write(status) == (glob_stee_sndary_status' = fetch_loc(status))
                   /\ fix_lvars /\ UNCHANGED(glob_stee_primary_status)

\*Main
main1 == n = "n_main1" /\ n' = "n_main2" /\ c_havoc /\ fix_stack /\ fix_lvars
main2 == n = "n_main2" /\ n' = "n_main4" /\ fix_stack /\ fix_lvars /\ 1 = 0
main3 == n = "n_main2" /\ n' = "n_stee1"  /\
         push("n_main3", lvars) /\ init_stee(lvars') /\ fix_globs /\ 1 # 0
main4 == n = "n_main3" /\ n' = "n_main1" /\ fix_stack /\ fix_lvars
         /\ fix_globs
main5 == n = "n_main4" /\ n' = "n_main4" /\ fix_stack /\ fix_lvars
         /\ fix_globs

\*Stee
stee1 == n = "n_stee1" /\ n' = "n_stee2" /\ fix_stack /\ c_read(primary_info)
stee2 == n = "n_stee2" /\ n' = "n_stee3" /\ fix_stack
         /\ c_eval(primary_info, sndary_info)
stee3 == n = "n_stee3" /\ n' = "n_stee4" /\ fix_stack /\ c_write(sndary_info)
stee4 == n = "n_stee4" /\ pop(n', lvars') /\ fix_globs

\*Program
Init == n = "n_main1" /\ init_main(lvars) /\ init_stack /\ init_globals
M == (main1 \/ main2 \/ main3 \/ main4 \/ main5 \/
      stee1 \/ stee2 \/ stee3 \/ stee4)

Spec == Init /\ [][M]_all_vars
==================================================================
\end{lstlisting}
´
\subsection{\nuxmv{} model}
The \nuxmv{} model is similar to the $\tlaplus$ one, but differs slightly as \nuxmv{} is, for example, more strictly types than $\tlaplus$. Since array sizes must be fixed to a constant in \nuxmv{}, we set the stack to be of size 10, as we know this to be enough in this model. 
\begin{lstlisting}[language=TLA]
MODULE main

VAR
    nstack : array 0..1 of {n_main1, n_main2, n_main3, n_main4, n_main5,
                            n_stee1, n_stee2, n_stee3, n_stee4,dummy};
    vstack : array 0..1 of array 0..1 of 0..10;
    curr_top : 0..1;
    --\* Global vars
    glob_stee_primary_status : 0..10;
    glob_stee_sndary_status : 0..10;
    --\* Current control state and local vars
    n : {n_main1, n_main2, n_main3, n_main4, n_main5, n_stee1, n_stee2, n_stee3, n_stee4};
    lvars : array 0..1 of 0..10;


DEFINE
  --loc vars index
  primary_info := 0;
  sndary_info := 1;

--\*Inits
  init_main_lvars := [0,0];
  init_stee_lvars := next(lvars[primary_info]) = 0 & next(lvars[sndary_info]) = 0;

  init_globals := glob_stee_primary_status in 0..1 &
                  glob_stee_sndary_status in 0..1;
  init_vstack := [[0,0],[0,0]];
  init_nstack := [dummy,dummy];



-- Auxiliary funcs
  pop := next(lvars[0]) = vstack[curr_top][0]
         & next(lvars[1]) = vstack[curr_top][1] & next(n) = nstack[curr_top]
         & next(curr_top) = curr_top - 1;
  push := next(vstack[curr_top][0]) = lvars[0]
          & next(vstack[curr_top][1]) = lvars[1]
          & next(nstack[curr_top]) = n & next(curr_top) = curr_top + 1;
  fix_globs := next(glob_stee_sndary_status) = glob_stee_sndary_status
               & next(glob_stee_primary_status) = glob_stee_primary_status;
  fix_lvars := next(lvars[0]) = lvars[0] & next(lvars[1]) = lvars[1];
  fix_stack := next(vstack[0][0]) = vstack[0][0]
               & next(vstack[0][1]) = vstack[0][1]
               & next(vstack[1][0]) = vstack[1][0]
               & next(vstack[1][1]) = vstack[1][1];

-- contracts
  c_havoc := next(glob_stee_primary_status) in 0..10
             & next(glob_stee_sndary_status) = glob_stee_sndary_status;
  c_read := next(lvars[primary_info]) = glob_stee_primary_status
            & fix_globs;
  c_eval := (lvars[primary_info] != 0 -> next(lvars[sndary_info]) = 0)
            & (lvars[primary_info] = 0 -> next(lvars[sndary_info]) = 1)
            & fix_globs;
  c_write := (next(glob_stee_sndary_status) = lvars[sndary_info])
             & fix_lvars
             & next(glob_stee_primary_status) = glob_stee_primary_status;


-- main
  Main1 := (n = n_main1 & next(n) = n_main2 & c_havoc & fix_lvars & fix_stack);
  Main2 := (n = n_main2 & next(n) = n_main4 & fix_lvars & fix_stack & 1 = 0);
  Main3 := n = n_main2 & next(n) = n_main3 & fix_stack & fix_lvars & push
           & init_stee_lvars & 1 = 0;
  Main4 := n = n_main3 & next(n) = n_stee1 & fix_stack & fix_lvars & fix_globs;
  Main5 := n = n_main4 & next(n) = n_main4 & fix_stack & fix_lvars & fix_globs;

-- stee
  Stee1 := n = n_stee1 & next(n) = n_stee2 & fix_stack & c_read;
  Stee2 := n = n_stee2 & next(n) = n_stee3 & fix_stack & c_eval;
  Stee3 := n = n_stee3 & next(n) = n_stee4 & fix_stack & c_write;
  Stee4 := n = n_stee4 & pop & fix_globs;

INIT
  lvars = init_main_lvars & vstack = init_vstack & curr_top = 0 &
  nstack = init_nstack & init_globals & n = n_main1;

TRANS
  Main1 | Main2 | Main3 | Main4 | Main5 | Stee1 | Stee2 | Stee3 | Stee4;
\end{lstlisting}

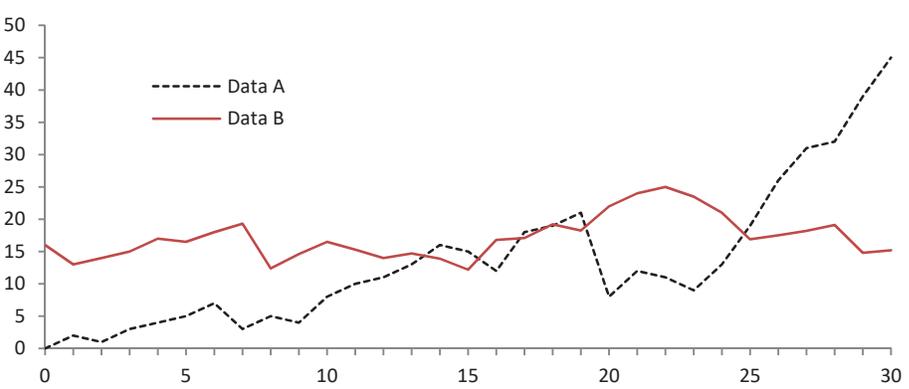
\end{document}